\documentclass[aps,prb,reprint,floatfix,nobibnotes]{revtex4-1}

\usepackage{mathtools,amssymb}
\usepackage{units,siunitx}
\usepackage{graphicx}
\usepackage[colorlinks,hyperindex]{hyperref}
\hypersetup{colorlinks,citecolor=blue,linkcolor=blue,urlcolor=blue}
\usepackage[version=4]{mhchem}

\begin{document}

\title{Role of Oxygen Adsorption in Nanocrystalline ZnO Interfacial Layers for Polymer--Fullerene Bulk Heterojunction Solar Cells}

\author{Sebastian Wilken}
\email{sebastian.wilken@uol.de}
\author{J\"urgen Parisi}
\author{Holger Borchert}

\affiliation{Institute of Physics, Energy and Semiconductor Research Laboratory, Carl von Ossietzky University of Oldenburg, 26111 Oldenburg, Germany}

\begin{abstract}
Colloidal zinc oxide~(ZnO) nanoparticles are frequently used in the field of organic photovoltaics for the realization of solution-producible, electron-selective interfacial layers. Despite of the widespread use, there is still lack of detailed investigations regarding the impact of structural properties of the ZnO particles, like the particle size and shape on the device performance. In the present work, ZnO nanoparticles with varying surface-area-to-volume ratio were synthesized and implemented into indium tin oxide-free polymer--fullerene bulk heterojunction solar cells featuring a gas-permeable top electrode. By comparing the electrical characteristics before and after encapsulation from the ambient atmosphere, it was found that the internal surface area of the ZnO layer plays a crucial role under conditions where oxygen can penetrate into the solar cells. The adsorption of oxygen species at the nanoparticle surface is believed to cause band bending and electron depletion next to the surface. Both effects result in the formation of a barrier for electron injection and extraction at the ZnO/bulk heterojunction interface and were more pronounced in case of small ZnO nanocrystals with a high surface-area-to-volume ratio. Different transport-related phenomena in the presence of oxygen are discussed in detail, i.e., increasing Ohmic losses, expressed in terms of series resistance, as well as the occurrence of space-charge-limited currents, related to the accumulation of charges in the polymer--fullerene blend. Since absorption of UV light can cause desorption of adsorbed oxygen species, the electrical properties depend also on the illumination conditions. With the help of systematic investigations of the current versus voltage characteristics of solar cells under different air exposure and illumination conditions as well as studies of the photoconductivity of pure ZnO nanoparticle layers, we gain detailed insight into the role of the ZnO nanoparticle surface for the functionality of the organic solar cells.
\end{abstract}

\maketitle


\section{Introduction}
In polymer-based bulk heterojunction solar cells,\cite{Deibel2010} the absorber blend, intrinsically, has no preferential transport direction for photogenerated charge carriers due to the statistical intermixing of both the donor and acceptor phase. Therefore, charge-selective interfacial layers,\cite{Steim2010,Ratcliff2011} which are semipermeable membranes for either electrons or holes in an ideal case,\cite{Wuerfel2009} are widely applicated to achieve efficient charge extraction at the respective electrodes and determine the device polarity. Amongst the materials for electron collection and transport, ZnO nanoparticles have attracted a lot of attention in polymer photovoltaics,\cite{Hau2008,Huang2011,You2013} as they combine the favorable properties of crystalline ZnO~(e.g., high electron mobility, wide band gap, n-type conductivity) with the low temperature and solution-based processing methods of colloidal chemistry.\cite{Pacholski2002,Beek2005} The most critical issue related to thin films made from colloidal nanoparticles is their large internal surface area. This typically results in a high density of surface defects, and the recombination of excess charge carriers at the ZnO/bulk heterojunction interface should be considered as potential loss mechanism in polymer solar cells comprising nanocrystalline ZnO interlayers.\cite{Chen2012,Wu2013,Shao2013}

\begin{figure}
\includegraphics{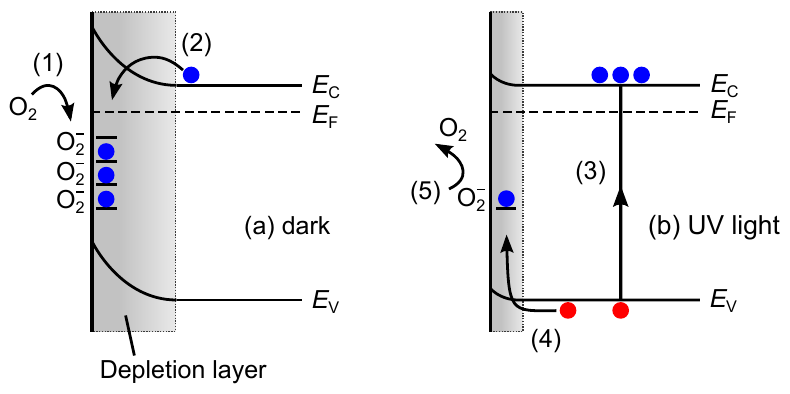}%
\caption{Schematic illustration of the oxygen adsorption--desorption process in n-type ZnO nanostructures. (a)~In the dark, chemisorption of \ce{O2} molecules~(1) via capturing free electrons from the conduction band~(2) leads to the formation of a depletion layer next to the surface. (b)~Under UV illumination, electron-hole pairs are generated~(3), and free holes likely recombine with the trapped electrons~(4). Consequently, the $\text{O}_2^-$ species are neutralized and become detached from the surface~(5).}%
\label{fig:adsorption_desorption}
\end{figure}

Another well-known feature of ZnO surfaces is the sensitivity to several adsorbates. One prominent example is the chemisorption of ambient oxygen, schematically illustrated in Figure~\ref{fig:adsorption_desorption}. It is generally accepted that chemisorbed oxygen species~(e.g., \ce{O2} molecules) are able to extract electrons from the conduction band of ZnO~($\text{O}_2(\text{g}) + \text{e}^- \rightarrow \text{O}_2^-(\text{ad})$). Consequently, an electron depletion layer forms in the vicinity of the surface, resulting in upward band bending. The vice versa process has been observed under the presence of reducing species, e.g., atomic hydrogen,\cite{Li2012} which would rather lead to electron accumulation and downward band bending at the surface. However, upon illumination with a photon energy above the band gap energy~(UV light), electron--hole pairs are created, and photogenerated holes are likely to migrate to the surface and recombine with trapped electrons. Accordingly, the chemisorbed $\text{O}_2^-$ species become discharged and detach from the surface~($\text{O}_2^-(\text{ad}) + \text{h}^+ \rightarrow \text{O}_2(\text{g})$), resulting in a reduction of the depletion layer width. It is noteworthy that a certain probability for re-adsorption exists, and the resulting adsorption--desorption balance strongly depends on the present oxygen pressure and UV illumination intensity. As being a surface-related phenomenon, the oxygen adsorption--desorption mechanism is particularly pronounced in nanostructures with high surface-area-to-volume ratio, such as nanowires and colloidal nanoparticles, which are, thus, intensively studied for gas sensing,\cite{Ghosh2011,Gurwitz2014} UV detection,\cite{Jin2008,Dhara2011,Guo2012} and photocatalytic applications.\cite{Bora2013}

Despite the widespread use of ZnO nanoparticles in polymer solar cells, only few publications discuss the influence of ambient oxygen on the electrical characteristics in detail. Lilliedal et al.\cite{Lilliedal2010} investigated flexible roll-to-roll processed devices, comprising a ZnO layer at the bottom electrode. The fabrication was performed in ambient air, and the current versus voltage characteristics of fresh devices exhibited a strong inflection point or S-shaped behavior, which was found to disappear after a certain photo-annealing procedure~(i.e., illumination at elevated temperatures). In that context, also a reverse voltage bias treatment~(\unit[$-20$]{V}) and subsequent light-soaking was discussed.\cite{Manor2012} Recently, Morfa et al.\cite{Morfa2014} studied the UV-enhanced conductivity of nanocrystalline ZnO interlayers in inverted polymer solar cells under consideration of surface photochemistry and regarding the impact of molecular oxygen and water. To date, most literature studies focus on devices with opaque metallic top electrodes, whereas the latter can serve as diffusion barrier for ambient gases. From a more commercial point of view, fully solution-processed polymer solar cells are of great interest, and high-conductivity poly(3,4-ethylenedioxythiophene):poly(styrenesulfonate)~(PEDOT:PSS) is frequently used for the realization of printable top electrodes.\cite{Gaynor2010,Zhou2010} However, polymer films like those made from PEDOT:PSS are known to show a distinct gas permeability,\cite{Andersen2007,Seemann2009,Norrman2010,Voroshazi2011} and, hence, fully solution-processed devices containing ZnO nanoparticles are supposed to be particularly sensitive against ambient oxygen.

In this study, we investigate the influence of the ambient atmosphere on the functionality of nanocrystalline ZnO interfacial layers, which were implemented into inverted poly(3-hexyl\-thiophene)~(P3HT):[6,6]-phenyl-C$_{61}$-butyric acid methyl ester~(P3HT:PCBM) bulk heterojunction solar cells in substrate configuration,\cite{Glatthaar2005,Wilken2012a} comprising an opaque Cr/Al/Cr back contact and a gas-permeable PEDOT:PSS top electrode. Previously, we have reported that the introduction of an interfacial ZnO layer can lead to significantly improved electron extraction at the metallic bottom electrode.\cite{Wilken2012b} Here, we demonstrate that without a suitable encapsulation both the electron injection and extraction at the ZnO/bulk-heterojunction interface are drastically degenerating upon exposure to ambient air on a relatively short time scale and can only partly be recovered using UV containing illumination~(light-soaking). By using ZnO nanoparticles with varying surface-area-to-volume ratio, we show that charge injection/extraction is highly sensitive to the internal surface area of the nanocrystalline ZnO film. Depending on the nanoparticle size, different transport-related phenomena are discussed in detail, i.e., Ohmic losses, expressed in terms of an increasing series resistance, as well as the occurrence of space-charge-limited currents, related to the formation of an extraction barrier at the ZnO/bulk-heterojunction interface. 


\section{Experimental Methods}
\subsection{Synthesis of ZnO Nanoparticles}
Zinc oxide nanoparticles were synthesized from zinc acetate dihydrate~(\ce{ZnAc2*2H2O}, Sigma Aldrich, 98\%) and potassium hydroxide~(\ce{KOH}, Carl Roth, 85\%) in methanol, as described in more detail elsewhere.\cite{Wilken2012b,Pacholski2002} A variation in particle size and shape has been achieved by changing the amount of precursors, while all other synthesis parameters, in particular the ratio between \ce{ZnAc2*2H2O} and \ce{KOH}, were kept constant. The as-obtained particles were precipitated, washed twice with methanol, and re-dispersed without any additional surfactants with high concentration~(\unit[50 to 250]{mg/ml}) in a binary mixture of chloroform and methanol~(9:1 v:v). Prior application, stock dispersions were further diluted to yield the desired concentration and filtered through a \unit[0.2]{$\mu$m} syringe filter.

\subsection{Solar Cell Fabrication}
For the inverted solar cells, the bottom contact was thermally evaporated through a shadow mask~($\unit[1 \times 1]{cm^2}$) on cleaned glass substrates in sequence Cr~(\unit[6]{nm}), Al~(\unit[100]{nm}), and Cr~(\unit[12]{nm}). Subsequently, the ZnO nanoparticles~(\unit[5]{mg/ml}) were spin-coated at \unit[1500]{rpm} in a nitrogen-filled glove box~($\ce{O2} < \unit[1]{ppm}$). Without further treatment of the ZnO layer, a \unit[120]{nm} thick 1:1 wt:wt absorber blend of P3HT~(Merck KGaA, 94.2\% regioregularity, Mw = \unit[54\,200]{g/mol}, Mn = \unit[23\,600]{g/mol}) and PCBM~(Solenne BV, 99.5\%) was then spun from 1,2-dichlorobenzene, slowly dried over  about \unit[1]{h} in a covered petri dish, and afterwards annealed at \unit[150]{$^\circ$C}. High-conductivity PEDOT:PSS~(Heraeus Clevios CPP 105 D) was spin-coated on top of the P3HT:PCBM blend and annealed at \unit[110]{$^\circ$C}. Finally, a current-collecting Ag grid with a grating period of \unit[1.1]{mm} was thermally evaporated under high vacuum~(\unit[$10^{-6}$]{mbar}). For encapsulation, devices were sealed with an UV curable optical adhesive~(DYMAX Light Weld) and glass slides. The adhesive was cured using a  UV lamp~(\unit[350]{nm}, \unit[1]{mW/cm$^2$}). Encapsulation was done in ambient air.

\subsection{Characterization Techniques}
Structural, morphological, and optical properties of the ZnO nanoparticles were obtained using transmission electron microscopy~(Zeiss EM 902A), X-ray diffraction~(PANalytical X'PertPro MPD), atomic force microscopy~(Agilent 5420), UV-visible absorption~(Varian Cary 100), and photoluminescence spectroscopy~(Horiba FluoroLog-3). The optical measurements were performed on solutions with low concentration using quartz cuvettes. Background signals resulting from pure solvents were subtracted if necessary. Current density versus voltage characteristics of the inverted solar cells were recorded, under nitrogen atmosphere, with a source-measuring unit~(Keithley 2400) and, in ambient air, with a semiconductor characterization system~(Keithley 4200). All measurements were performed using four-terminal sensing to avoid the influence of contact and wiring resistances. For illumination, a class AAA solar simulator~(Photo Emission Tech.) was used, providing a simulated AM1.5G spectrum. Light intensity was adjusted to \unit[100]{mW/cm$^2$} using a calibrated Si reference solar cell. The photoactive area of the solar cells~(\unit[1]{cm$^2$}) was delimited by the structured Cr/Al/Cr back electrode. Additionally, we calculated the spectral mismatch factor,\cite{snaith2012} which was found to be close to unity~(1.047) and, therefore, not further taken into account. 

\subsection{Charge Transport in ZnO Nanoparticle Films}
To study the charge transport properties of pure nanocrystalline ZnO films, we implemented them into a lateral metal--semiconductor--metal structure. Therefore, the ZnO nanoparticles were spin-coated on glass and annealed at \unit[150]{$^\circ$C} for \unit[10]{min} in a nitrogen-filled glove box. Subsequently, two planar Al electrodes with a thickness of  \unit[100]{nm} and a lateral spacing of \unit[70]{$\mu$m} were thermally evaporated under high vacuum~(\unit[$10^{-6}$]{mbar}) on top of the film. Current versus voltage characteristics of the Al/ZnO/Al devices were recorded in the dark and under UV illumination~(\unit[350]{nm}, \unit[1]{mW/cm$^2$}) using a source-measuring unit~(Keithley 2400). Measurements were performed both under nitrogen atmosphere and in ambient air.


\section{Results and Discussion}
\subsection{Structural and Optical Properties of the ZnO Particles}

\begin{figure}
\includegraphics{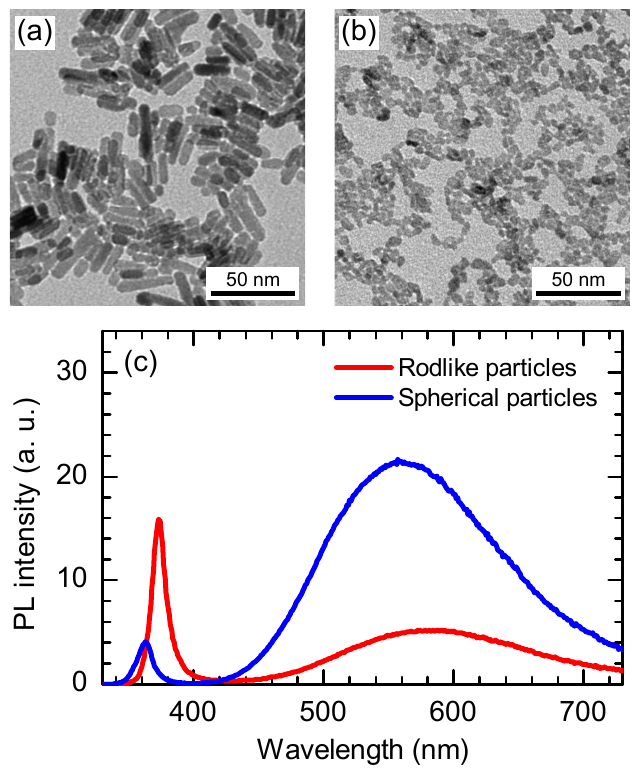}%
\caption{TEM images of the (a)~rod-like and (b)~spherically shaped ZnO nanoparticles used in this study. (c) Room-temperature PL spectra for both types of particles upon excitation at \unit[320]{nm}. For better comparison of the PL measurements, nanoparticle dispersions with similar optical density at the excitation wavelength have been used.}%
\label{fig:TEM_PL}
\end{figure}

Figures \ref{fig:TEM_PL}a and \ref{fig:TEM_PL}b show transmission electron microscopy~(TEM) images of the different types of ZnO nanoparticles discussed in this study, which were obtained using identical synthesis conditions, except the actual precursor concentration. It is well established for the method used that quasi-spherical ZnO nanoparticles can form nanorods by a growth mechanism called oriented attachment, if the precursor concentration is high during synthesis.\cite{Pacholski2002} The particles shown in Figure~\ref{fig:TEM_PL}a were prepared with a \ce{ZnAc2\cdot2 H2O} concentration of \unit[0.7]{mol/l}, which yielded nanorods. These particles will be referred to as ``rod-like particles'' in the following. With a five times lower concentration~(\unit[0.14]{mol/l}), the nanoparticles shown in Figure~\ref{fig:TEM_PL}b were obtained. In the TEM image, individual quasi-spherical particles as well as short rods, in most cases composed of only two attached ZnO particles, can be observed. For simplicity, the particles of this sample will be called ``spherical particles'' in the following, although the sample contains also a certain amount of short rods.

Powder X-ray diffraction~(XRD) measurements~(see Supplemental Material, Figure~S1) confirm the formation of wurtzite-type ZnO in both cases, with the spherical particles showing an increased peak broadening due to the smaller diffracting domains. In order to obtain statistical information about the average crystallite size and shape, the measured XRD patterns were modeled by means of Rietveld refinement using the software MAUD,\cite{Lutterotti2004} which can take into account the anisotropic crystallite shape and texture effects. From that analysis, we derived an average crystallite size of \unit[$27 \times 8.8$]{nm}~(rod-like particles) and \unit[$6.8 \times 4.5$]{nm}~(spherical particles). Thus, in average, the rod-like particles~(aspect ratio 3:1) consist of three attached ZnO particles, whereas the other sample~(aspect ratio 1.5:1) may be considered as a mixture of spherical particles and short rods, composed of two crystallites. Assuming sphero-capped cylinders, this corresponds to a geometric surface-area-to-volume ratio of \unit[0.51]{nm$^{-1}$}~(rod-like particles) and \unit[1.14]{nm$^{-1}$}~(spherical particles), respectively.

Figure~\ref{fig:TEM_PL}c shows room temperature photoluminescence~(PL) spectra of the ZnO nanoparticles excited at \unit[320]{nm}~(\unit[3.88]{eV}). The PL spectra exhibit two emission bands of varying intensity, a characteristic near band-edge emission in the UV, related to band--band transitions, and a broad emission in the visible range, which is often referred to as green luminescence and  attributed to trap-assisted recombination. The observed values of the peak wavelength are \unit[373]{nm}~(\unit[3.32]{eV}) and \unit[590]{nm}~(\unit[2.2]{eV}) for the rod-like particles, and \unit[363]{nm}~(\unit[3.42]{eV}) and \unit[565]{nm}~(\unit[2.1]{eV}) for the spherical particles, respectively. Due to quantum confinement, both emission features are blue-shifted in case of the smaller spherical particles relative to the rod-like ones, which is in agreement with the variation of the optical band gap energy from \unit[3.34]{eV}~(rod-like) to \unit[3.43]{eV}~(spherical), as estimated by the absorption onset using the Tauc procedure for direct semiconductors~(see Supplemental Material, Figure~S2). However, more significant here is the relative intensity between both emission features, and it can be seen that the visible luminescence becomes more pronounced in case of the smaller spherical particles compared to the larger rod-like ones, whereas the opposite trend is obvious for the UV emission band. The corresponding values of the intensity ratio~(visible to band edge emission), obtained from integrated peak areas with the abscissa converted to an energy scale, are $\sim 1.4$~(rod-like) and $\sim 21$~(spherical), respectively.

The visible luminescence in ZnO is commonly attributed to recombination of mobile~(or shallowly trapped) electrons with deeply trapped holes, and charged oxygen vacancies~(V$_\text{O}$) are often supposed as the corresponding recombination centers.\cite{vanDijken2004,Gong2007} However, first-principle calculations of the native point defects in ZnO showed that oxygen vacancies are deep donor states with a rather high formation energy, typically resulting in low V$_\text{O}$ concentrations in n-type ZnO.\cite{Janotti2007} Instead, zinc vacancies~(V$_\text{Zn}$) were suggested to be more likely as origin for the visible luminescence, which has also been confirmed experimentally.\cite{Tuomisto2003,Zhao2005}. It could further be shown that the intensity of the visible luminescence increases by the presence of chemisorbed oxygen species, such as $\text{O}_2^-$ and $\text{OH}^-$,\cite{Norberg2005,Ghosh2011} whereas it is effectively quenched after hydrogen annealing\cite{Li2012,Cooper2012} or by suitable surface modification with organic molecules.\cite{Gong2007,Norberg2005} Consequently, it appears reasonable to suggest that the amount of surface states significantly influences the population dynamics of the green luminescence centers~(e.g., V$_\text{Zn}$), as is also supported by our PL measurements on particles with varying surface-area-to-volume ratio. However, direct recombination via surface states is typically non-radiative, and, hence, the recombination of surface-trapped electrons and holes with their mobile counterparts~(which are still existing in the transport bands after photoexcitation) is not likely to explain the increase of the green luminescence with decreasing particle size. One possible explanation has been given in the work of van Dijken et al.\cite{vanDijken2004} Herein, surface-trapped holes in a surface system such as chemisorbed $\text{O}_2^-$ were considered, and the surface-trapped holes have been supposed to tunnel back into the particle where they give rise to the creation of recombination centers for the visible luminescence. The importance of surface hole-trapping was also highlighted in other studies.\cite{Li2012,Teklemichael2011} Nevertheless, the defect chemistry of ZnO and the origin of the green luminescence are still under debate.\cite{Zhang2010,Gurwitz2014,Cooper2012,Djurisic2006}

\begin{figure}
\includegraphics{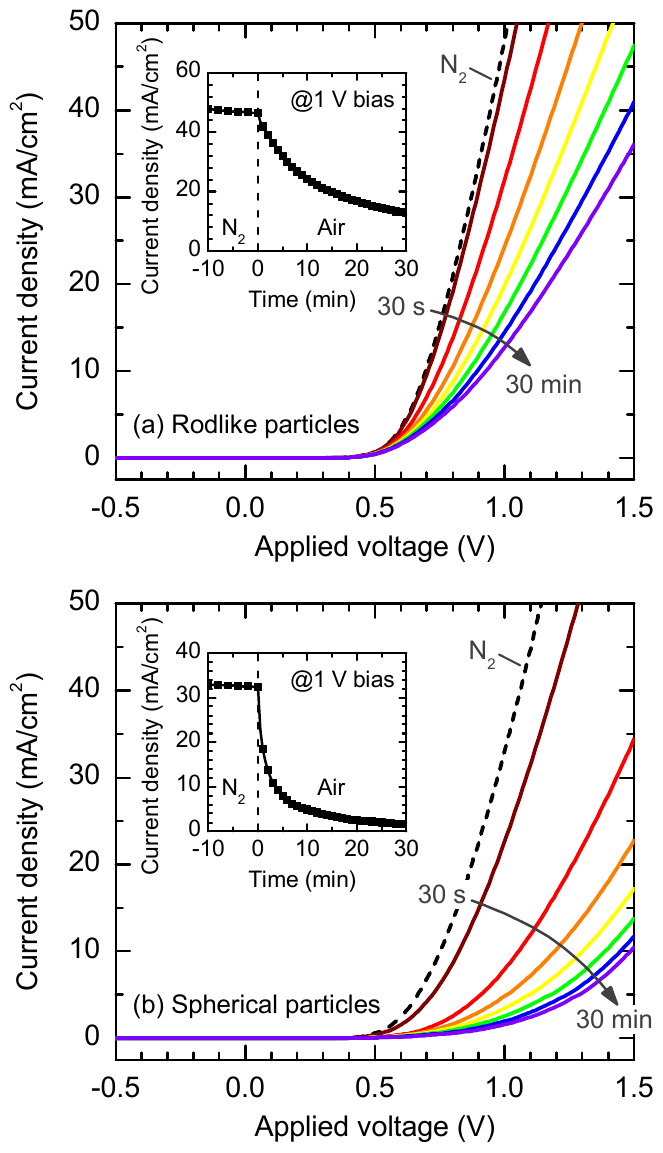}%
\caption{Dark $J$--$V$ characteristics of inverted P3HT:PCBM solar cells~(non-encapsulated) with interfacial layers made from the (a)~rod-like and (b)~spherical ZnO nanoparticles and comprising a gas-permeable PEDOT:PSS top electrode. Measurements were initially performed under nitrogen atmosphere~(dashed lines) and, subsequently, in ambient air~(solid lines) for different exposure times ranging from \unit[30]{s} to \unit[30]{min}~(after first air contact). Insets: Time development of the current density at \unit[+1]{V} bias under nitrogen atmosphere and in ambient air. The moment of the transfer to ambient air is taken as $t = 0$.}%
\label{fig:IV_dark}
\end{figure}

\subsection{Solar Cell Characteristics Depending on Atmosphere}
To examine the influence of ambient air on the electrical characteristics of polymer--fullerene solar cells containing nanocrystalline ZnO, we implemented thin films made from both types of ZnO particles as electron-selective interfacial layers into indium tin oxide-free inverted devices with the layer sequence Cr\slash{}Al\slash{}Cr\slash{}ZnO\slash{}P3HT:PCBM\slash{}PEDOT:PSS\slash{}Ag grid. High-conductivity grade PEDOT:PSS served as transparent top electrode, whereas a thermally evaporated Ag grid was additionally employed to enhance the lateral conductivity. Details on the device architecture can be found elsewhere.\cite{Wilken2012a,Wilken2012b} The ZnO nanoparticles were applied via spin-coating on the metallic back electrode without any subsequent treatment, and the film thickness was about \unit[20]{nm} for both the rod-like and spherical particles, as determined by means of spectral ellipsometry~(not shown here). We further investigated the surface morphology using atomic force microscopy~(see Supplemental Material, Figure~S3) and the root-mean-square roughness of films made from the rod-like particles~(\unit[11.7]{nm}) is found to be larger as in case of the spherical particles~(\unit[3.2]{nm}). This appears reasonable concerning the larger size and the anisotropic shape of the rod-like particles, resulting in a more nonuniform packing in the deposited nanocrystalline films.

Figure~\ref{fig:IV_dark} shows dark current density versus voltage~($J$--$V$) characteristics of representative~(non-encapsulated) inverted solar cells recorded under different gas environments. After fabrication, the samples were kept under nitrogen atmosphere and the $J$--$V$ characteristics initially measured over a period of \unit[10]{min}. Only negligible changes in the electrical behavior under applied forward bias could be observed during storage under nitrogen, as is visualized by the time development of the current density at a bias voltage of \unit[+1]{V}~(insets). Subsequently, the solar cells were transferred to ambient air, and the $J$--$V$ characteristics were recorded iteratively between \unit[30]{s} and \unit[30]{min} after first contact to air. For both types of nanoparticles, the forward current then exhibits a gradual decrease in magnitude on a relatively short time scale, indicating the degradation of the charge carrier injection and/or transport properties of~(at least) one of the electrodes. Apparently, the decay is much faster in case of the smaller, spherically shaped particles. After \unit[30]{min} of air exposure, a reduction of 73\%~(95\%) in the forward current density~(at \unit[+1]{V} bias) can be seen for the rod-like~(spherical) nanoparticles compared to the initial state under nitrogen atmosphere. Also changes of the shape of the $J$--$V$ curves can be observed upon longer air exposure, especially pronounced for the spherical particles, which further points to a change of the dominant transport mechanism. The latter will be discussed in more detail below. Hence, the different behavior with varying nanoparticle size, whereas all other components of the solar cells were identical, strongly suggest the ZnO/bulk heterojunction interface as origin for the inferior charge carrier injection/transport properties in ambient air.

For an accurate investigation of the electrical characteristics under different illumination conditions, we fabricated a batch of 12 individual solar cells in total, 6 of each comprising interlayers made from the rod-like and the spherical particles, respectively. After fabrication under inert conditions, the devices were transferred to ambient air and kept in dark for a certain time. Subsequently, the $J$--$V$ characteristics were measured (i) in the dark, (ii) under AM1.5G illumination with a UV blocking filter~(long pass, \unit[400]{nm}), and (iii) under standard AM1.5G illumination~(\unit[100]{mW/cm$^2$}). The samples were then transferred back to the inert atmosphere and annealed at \unit[110]{$^\circ$C} for \unit[10]{min}, which was found to entirely recover the initial state. This cycle was performed for three different storage times in ambient air~(2, 5, and \unit[10]{min}) and for each device individually. Afterwards, all devices were encapsulated in ambient air using glass slides and an optically clear adhesive, which was cured using UV illumination. Finally, the $J$--$V$ characteristics of the encapsulated devices were recorded under the same illumination conditions as stated before.

\begin{figure}
\includegraphics{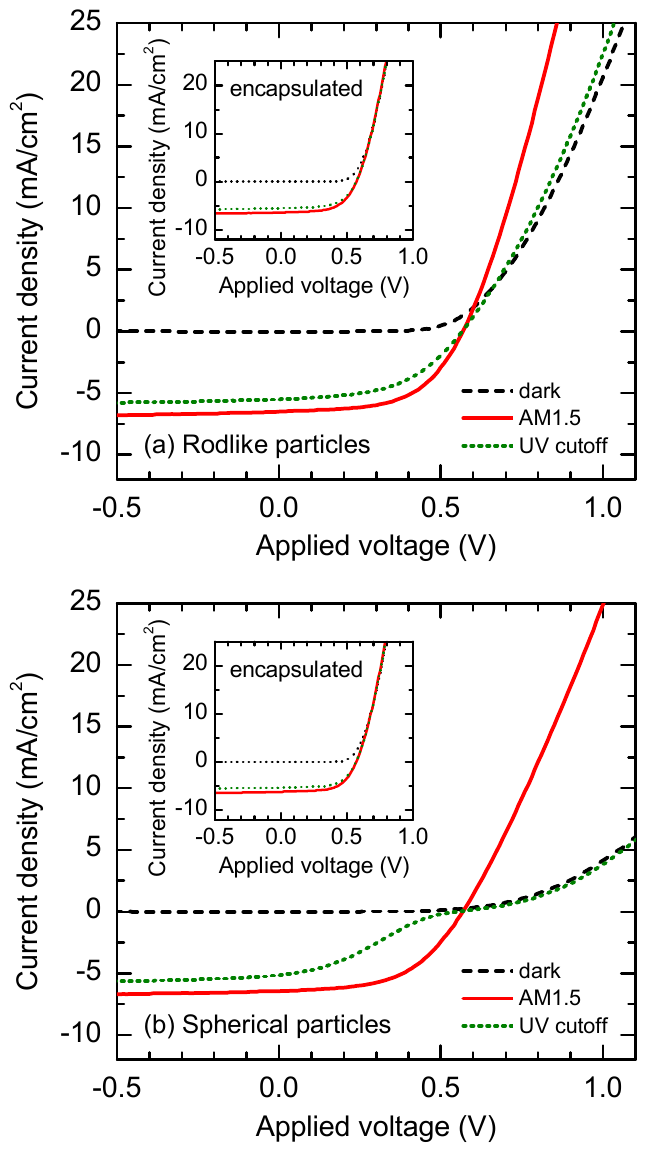}%
\caption{$J$--$V$ characteristics of inverted P3HT:PCBM solar cells~(non-encapsulated) with interfacial layers made from the (a)~rod-like and (b)~spherical ZnO nanoparticles after \unit[5]{min} storage in ambient air. Measurements were performed subsequently in the dark~(black dashed lines), under AM1.5G illumination with a \unit[400]{nm} long pass filter~(green dotted lines), and under standard AM1.5G illumination~(red solid line). Insets: Same measurement routine performed after encapsulation of the devices.}%
\label{fig:IV_AM1.5}
\end{figure}

\begin{table*}
\caption{Photovoltaic performance parameters of solar cells with interlayers made from the rod-like and spherical ZnO nanoparticles under simulated AM1.5G illumination.\protect\footnote{Devices were initially characterized without encapsulation after different storage time in ambient air~(2, 5, and \unit[10]{min}) and, subsequently, after having been encapsulated. The values reported are mean values averaged over 6 individual solar cells in each case and the corresponding standard deviation.}}
\label{tab:IV_Params}
\centering
\begin{ruledtabular}
\begin{tabular}{llccccc}
    Particles & Encapsulation & $V_\text{oc}$~(mV) & $J_\text{sc}$~(mA/cm$^{2}$) & FF~(\%) & PCE~(\%)  & $r_\text{oc}$~($\Omega$cm$^2$)\\
\noalign{\vskip .75mm}
\hline
\noalign{\vskip .75mm}
   	rod-like & without~(\unit[2]{min}) & 564 $\pm$ 4& 6.5 $\pm$ 0.3 & 56.4 $\pm$ 1.6 & 2.08 $\pm$ 0.12 & 17 $\pm$ 1\\
    & without~(\unit[5]{min}) & 567 $\pm$ 3 & 6.5 $\pm$ 0.4 & 56.0 $\pm$ 2.0 & 2.07 $\pm$ 0.14 & 18 $\pm$ 2\\
    & without~(\unit[10]{min}) & 569 $\pm$ 4 & 6.5 $\pm$ 0.4 & 54.6 $\pm$ 2.7 & 2.01 $\pm$ 0.15 & 21 $\pm$ 3\\
    & with & 570 $\pm$ 6 & 6.4 $\pm$ 0.4 & 59.0 $\pm$ 2.4 & 2.14 $\pm$ 0.13 & 16 $\pm$ 1\\[.5em]
    spherical & without~(\unit[2]{min}) & 563 $\pm$ 4 & 6.5 $\pm$ 0.3 & 54.0 $\pm$ 1.1 & 1.98 $\pm$ 0.13 & 21 $\pm$ 2\\
    & without~(\unit[5]{min}) & 567 $\pm$ 1 & 6.5 $\pm$ 0.2 & 52.2 $\pm$ 1.2 & 1.92 $\pm$ 0.11 & 24 $\pm$ 2\\
    & without~(\unit[10]{min}) & 569 $\pm$ 4 & 6.3 $\pm$ 0.2 & 49.2 $\pm$ 1.6 & 1.77 $\pm$ 0.10 & 31 $\pm$ 2\\
    & with & 570 $\pm$ 4 & 6.2 $\pm$ 0.2 & 60.8 $\pm$ 0.9 & 2.14 $\pm$ 0.10 & 16 $\pm$ 2\\
  \end{tabular}%
\end{ruledtabular}
\end{table*}

In Figure~\ref{fig:IV_AM1.5}, the $J$--$V$ characteristics obtained after \unit[5]{min} storage in ambient air are shown exemplarily for the devices without encapsulation, as well as after encapsulation~(insets). Complete data for all storage times can be found in the Supplemental Material~(Figures S4 and S5). The averaged device characteristics, i.e., the open circuit voltage~($V_\text{oc}$), the short circuit current density~($J_\text{sc}$), the fill factor~(FF), the power conversion efficiency~(PCE), and the open circuit differential resistance~($r_\text{oc} = \text{d}V/\text{d}J|_{J=0}$) are summarized in Table \ref{tab:IV_Params}. From Figure~\ref{fig:IV_AM1.5}, it can clearly be seen that the charge transport under forward voltage bias strongly depends on illumination conditions in case of the non-encapsulated devices. Under standard AM1.5G illumination, the forward current density significantly increases compared to the dark case, whereas it is rather unaffected when the UV part is removed from the illumination source~(i.e., when photoexcitation of the ZnO particles is suppressed).

As already shown for the dark case~(Figure~\ref{fig:IV_dark}), the effect of ambient air is more pronounced in case of the smaller spherical particles, resulting in an S-shaped $J$--$V$ curve under UV-filtered illumination~(Figure~\ref{fig:IV_AM1.5}b), whereas the S-shape disappears once the UV filter is removed. However, even in the presence of UV light, the values of the forward current density are lower than those in the initial dark $J$--$V$ measurements under nitrogen atmosphere, and the discrepancy increases with time, as is further displayed by the variation of $r_\text{oc}$. Consequently, we observed a decrease in FF from 56.4\%~(54.0\%) to 54.6\%~(49.2\%) for the rod-like~(spherical) particles after \unit[2]{min} and \unit[10]{min} of air exposure, respectively, resulting in a gradual reduction of the PCE from 2.08\%~(1.98\%) to 2.01\%~(1.77\%). Apart from that, relatively slight variations of $V_\text{oc}$ and $J_\text{sc}$ are observed, which is probably related to a degradation of the polymer--fullerene blend and is beyond the scope of this study.

In contrast, after encapsulation, the charge injection and transport properties were no longer influenced by the illumination conditions, as is indicated by the identical slope of the $J$--$V$ curves in the dark, as well as under UV-filtered and standard AM1.5G illumination~(see insets in Figure~\ref{fig:IV_AM1.5}). In that case, no more substantial differences can be seen between solar cells fabricated with the rod-like and spherical particles, respectively, and very similar photovoltaic performances with an FF of $\sim 60\%$ and a PCE of $2.14\%$ have been reached. We also studied the long-term stability of the encapsulated solar cells and found that the beneficial effect of the encapsulation remained for at least several months. For instance, after having been stored in the dark over a period of \unit[10]{months}, the devices exhibited approximately 95\% of their initial PCE~(obtained directly after the encapsulation procedure) without any significant change in the charge injection behavior~(see Supplemental Material, Figure~S6 and Table~S1).

\subsection{Charge Injection and Extraction at the ZnO/Bulk Heterojunction Interface}
\begin{figure}
\includegraphics{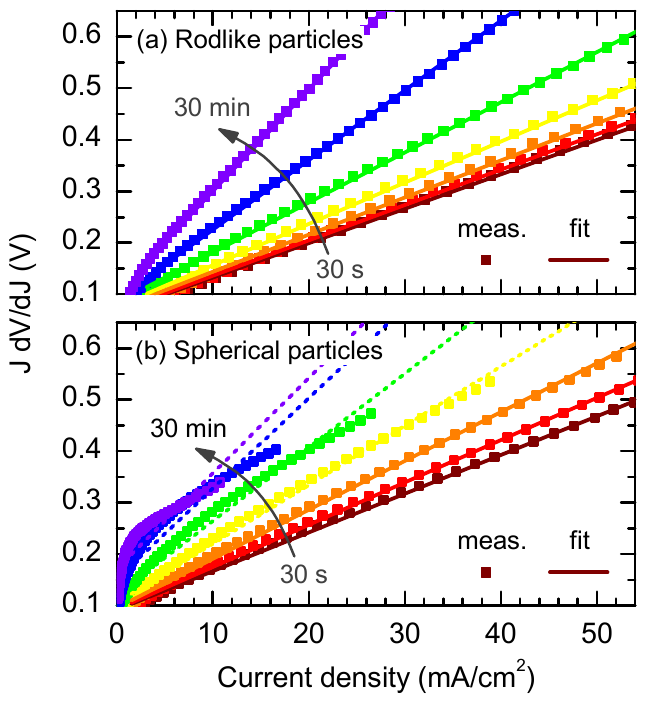}
\caption{Exemplary $J \text{d}V/\text{d}J$ versus $J$ representations of the dark $J$-$V$ characteristics under forward bias for different storage times in ambient air~(0.5, 1, 2, 4, 8, 16, and \unit[30]{min}). Lines indicate linear fits to the experimental data. Strong deviations from linear behavior are observed in case of the smaller spherical particles for longer air exposure~(dashed lines), indicating that the degradation of the $J$--$V$ curves can not be attributed simply to increasing series resistance in that case.}%
\label{fig:Rs}
\end{figure}

First, we discuss whether the decreasing forward current density of non-encapsulated devices upon exposure to ambient air can simply be described by a time-dependent increase of the series resistance $R_{\text{s}}$. When the forward-biased current density of a diode is dominated by $R_{\text{s}}$, the inverse slope~($\text{d}V/\text{d}J$) of the $J$-$V$ curve can be written as\cite{Hegedus2004,Schroder1990}
\begin{equation}
\frac{\text{d}V}{\text{d}J} = R_{\text{s}} + \frac{A k_{\text{B}} T}{q} J^{-1},
\label{eq:Rs}
\end{equation}
where $A$ is the diode ideality factor, $k_{\text{B}}$ Boltzmann's constant, $T$ the absolute temperature, and $q$ the elementary charge. Thus, plotting the quantity $J \text{d}V/\text{d}J$ versus $J$ would yield a straight line with an intercept of $A k_\text{B} T/q$ and a slope $R_{\text{s}}$. Figure~\ref{fig:Rs} shows exemplary $J \text{d}V/\text{d}J$ versus $J$ representations of the dark $J$--$V$ characteristics under forward bias, depending on the exposure time to ambient air ranging from \unit[0.5]{min} to \unit[30]{min}. Furthermore, linear fits of the experimental data according to Equation~(\ref{eq:Rs}) are shown.

For the rod-like particles, a reasonable fit could be obtained over the whole time range under consideration. This indicates that the degradation of the $J$--$V$ curves is mainly related to Ohmic losses and can effectively be described in terms of increasing series resistance. From the slopes of the linear fits, we observed a nearly linear increase in time of $R_{\text{s}}$, ranging from $\unit[6.7]{\Omega cm^2}$~(\unit[0.5]{min} air exposure) to $\unit[19.8]{\Omega cm^2}$~(\unit[30]{min} air exposure). In contrast, in case of the smaller spherical particles, the $J \text{d}V/\text{d}J$ versus $J$ data show strong deviations from linear scaling, especially pronounced for longer storage in ambient air. Consequently, the degradation of the $J$--$V$ curves is unlikely to be attributed to Ohmic losses in that case, and the evaluation of $R_\text{s}$ would not lead to meaningful results. Hence, a change of the dominant charge transport mechanism is suggested for the smaller particles in ambient air, which is further consistent with the inferior $J$--$V$ performance compared to the larger particles, especially concerning the more pronounced reduction of FF and PCE, and will be addressed in the following.

\begin{figure}
\includegraphics{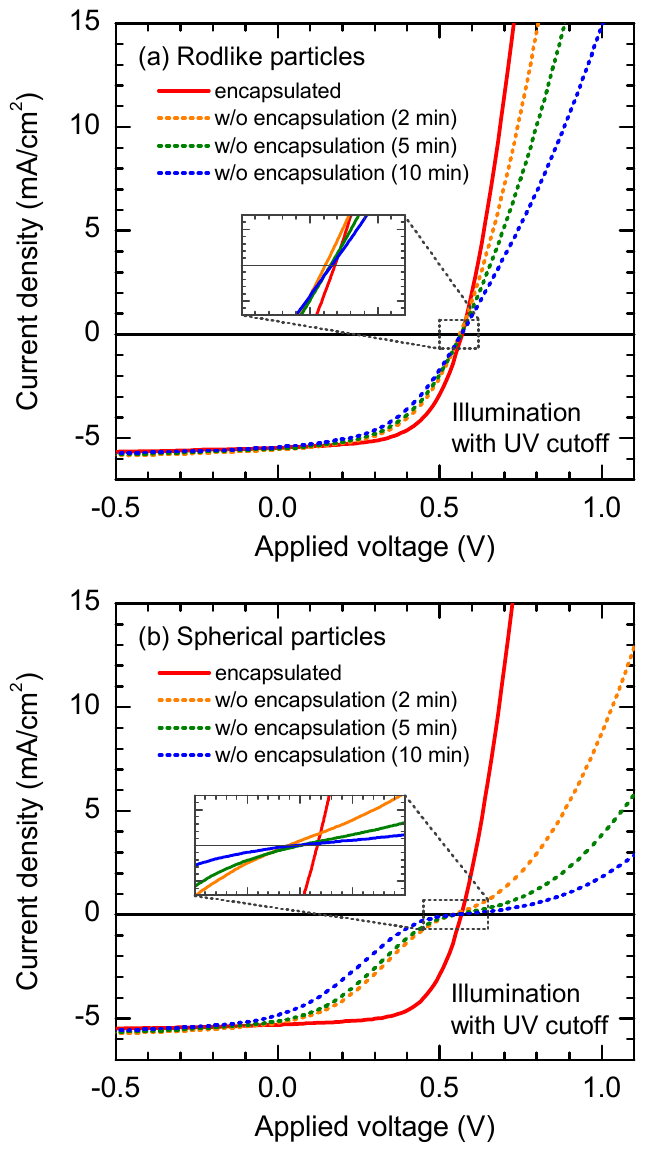}%
\caption{$J$--$V$ characteristics under UV cut-off AM1.5 illumination~(cut-off wavelength \unit[400]{nm}) for non-encapsulated~(dotted lines) and encapsulated~(solid lines) solar cells with interlayers processed from the (a)~rod-like and (b)~spherically shaped ZnO nanoparticles. For the non-encapsulated devices, the $J$--$V$ curves are shown for 2, 5, and \unit[10]{min} exposure time to ambient air.}%
\label{fig:SCLC_IV}
\end{figure}

To study the charge transport properties at the ZnO/bulk heterojunction interface in more detail, we focus now on the $J$--$V$ characteristics of the devices under UV-filtered illumination, as shown in Figure~\ref{fig:SCLC_IV}. The deformation of the $J$--$V$ curves in the forth quadrant can clearly be seen after storage in ambient air, in particular, the pronounced S-shape with an inflection point near the voltage axis intercept~($J = 0$) in case of the spherical particles. However, the value of $V_\text{oc}$ is only slightly affected  for both types of nanoparticles~($\Delta V_\text{oc} < \unit[20]{mV}$), whereas the voltage at the maximum power point~($V_\text{mpp}$) is shifted for non-encapsulated devices by $\Delta V_\text{mpp} = \unit[50]{mV}$~(rod-like particles) and $\Delta V_\text{mpp} = \unit[200]{mV}$~(spherical particles) after \unit[10]{min} storage in ambient air compared to the ideal case~(encapsulated).

An S-shaped deformation of the $J$--$V$ characteristics is often observed for organic solar cells with non-ideal contacts.\cite{Sandberg2014,Tress2011,Wagenpfahl2010} Recently, Sandberg et al.\cite{Sandberg2014} presented a comprehensive numerical study on several contact-related phenomena and discussed the effect of increased injection barriers, a reduced surface recombination velocity, interfacial minority carrier doping, and charge carrier trapping at the electrodes in detail. They also provide means to experimentally distinguish between the different mechanisms leading to S-shaped $J$-$V$ characteristics, and one criterion is the position of the inflection point. According to that analysis, an increasing injection barrier at the ZnO interface is not likely to solely explain the deformation of the $J$--$V$ curves in our case, as one would expect a distinct decrease of $V_\text{oc}$ related to the reduced built-in potential then. Also significant minority carrier doping or interfacial trapping appear unreasonable, as both should lead to an inflection point located well below the voltage axis~($J < 0$), and the forward current would not be decreasing in one of that cases. 

In a more favorable manner, our observations can be related to a reduced surface recombination velocity at the ZnO/bulk heterojunction interface, both limiting the current under extraction~($J < 0$) and injection~($J > 0$) conditions. Based on a numerical device model, Wagenpfahl et al.\cite{Wagenpfahl2010} showed that a reduction of the surface recombination velocity for electrons and holes~($S_\text{n,p}$) gives rise to an inflection point near $V_\text{oc}$, which they also confirmed experimentally by modifying the applied indium tin oxide anodes using different oxygen plasma treatments. $S_\text{n,p}$ quantitates the charge extraction for majority charge carriers at each electrode, and a reduction of $S_\text{n}$~($S_\text{p}$) points to an extraction barrier for electrons~(holes) at the cathode~(anode). Extraction barriers could arise due to an insulating layer adjacent to one of the electrodes or a large energetic barrier under charge extraction conditions.\cite{Tress2013} Tress et al.\cite{Tress2011} showed that an extraction barrier for photogenerated charge carriers shifts the onset of the forward current to higher voltages, whereas $V_\text{oc}$ remains nearly unaffected. This results in the formation of an inflection point near $V_\text{oc}$, and is in good agreement with our measurements.

\subsection{Unipolar Charge Transport in ZnO Nanoparticle Films}
\begin{figure}
\includegraphics{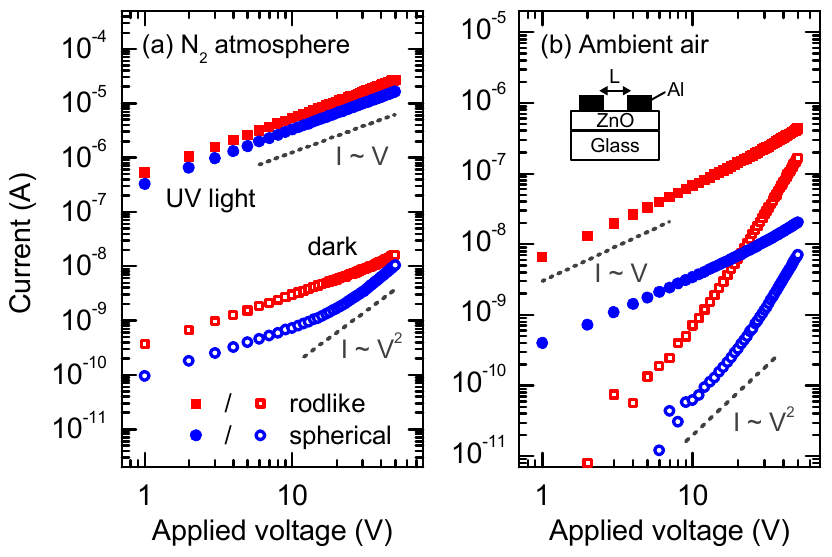}%
\caption{$I$--$V$ characteristics of ZnO thin films on glass with laterally spaced Al electrodes~(channel length $L = \unit[70]{\mu m}$) made from the rod-like~(squares) and spherical~(circles) particles, respectively, and recorded (a)~under nitrogen atmosphere and (b)~in ambient air. Measurements were both performed in the dark~(open symbols) and under \unit[350]{nm} UV illumination~(closed symbols). Dotted lines indicate the scaling expected for Ohmic~($I \propto V$) and trap-free SCLC~($I \propto V^2$) conduction. Inset: Schematic device architecture.}%
\label{fig:IV_NC}
\end{figure}
To further investigate the nature of the possible extraction barrier, we studied the unipolar charge transport properties of the ZnO nanoparticle films with respect to the surrounding ambient gas. For that purpose, we processed thin films from both types of ZnO nanoparticles on glass under the same conditions as for solar cell preparation and evaporated two planar Al electrodes on top, separated by a narrow gap~(length $L = \unit[70]{\mu m}$). Using that configuration, the conducting channel between the electrodes was in direct contact to the ambient gas, and current versus voltage~($I$--$V$) characteristics of the Al/ZnO/Al devices were recorded (i) under nitrogen atmosphere and (ii) in ambient air. The measurements were performed in the dark and upon exposure to UV illumination~(\unit[350]{nm}, \unit[1]{mW/cm$^2$}), respectively. Figure~\ref{fig:IV_NC} shows typical $I$-$V$ characteristics on a double-logarithmic scale, and the data points are analyzed in terms of power law relations, $I \propto V^\alpha$, where $\alpha$ is the scaling exponent. As ZnO/Al contacts are known to be Ohmic,\cite{Jin2008} no corrections have been made for the built-in voltage~($V_\text{bi} = \unit[0]{V}$). Under nitrogen atmosphere~(Figure~\ref{fig:IV_NC}a), we observed for both kinds of nanoparticles a linear scaling of the dark $I$-$V$ curve~($\alpha \sim 1$) at lower voltage, indicating Ohmic conduction. This implies that the density of mobile carriers within the nanoparticle films~($n$) dominates over the carrier density injected from the contacts~($n_\text{inj}$), and the conductivity $\sigma$ is given by
\begin{equation}
\sigma = q n \mu,
\end{equation} 
where $\mu$ is the charge carrier mobility. At larger bias voltages, a transition to higher scaling exponents can be seen. In case of the smaller spherical particles, the transition is more pronounced, and the scaling exponent reaches a value of $\alpha \sim 2$, being characteristic for the occurrence of space-charge-limited currents~(SCLC). In the SCLC regime, the current is dominated by injected charge carriers~($n_\text{inj} \gg n$). The current density can then be described by the Mott--Gurney law,
\begin{equation}
J = \frac{9 \varepsilon_r \varepsilon_0 \mu V^2}{8 L^3},
\end{equation}
where $\varepsilon_r$ denotes the relative permittivity, $\varepsilon_0$ the vacuum permittivity, and $L$ the length of a plane-parallel sample.

However, the current increases by several orders of magnitude upon UV illumination, and the photoconduction was found to be Ohmic over the whole range of the applied bias voltage for both types of nanoparticles. From the Ohmic regime, we calculated the conductivity, $\sigma = (J \cdot L)/(V)$, where $J$ denotes the current density normalized to the contact area, estimated from the product of the film thickness~(\unit[20]{nm}) and the contact width~(\unit[4]{mm}). From that analysis, we derived a dark conductivity of \unit[$2.3 \times 10^{-4}$]{S/m}~(\unit[$6.2 \times 10^{-5}$]{S/m}) and a photoconductivity of \unit[$4.5 \times 10^{-1}$]{S/m}~(\unit[$2.8 \times 10^{-1}$]{S/m}) for the rod-like~(spherical) particles, resulting in a photo-to-dark conductivity ratio of $\sim2\,000$~(rod-like) and $\sim4\,500$~(spherical), respectively.

\begin{figure}
\includegraphics{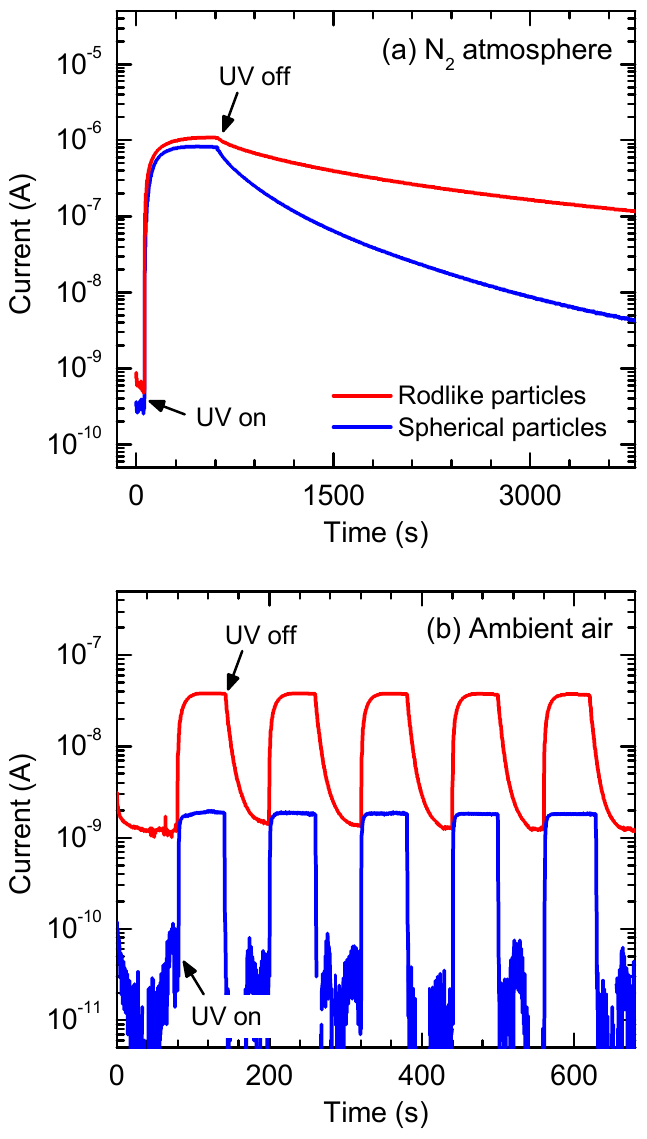}%
\caption{Transient photocurrent of the ZnO nanoparticles (a)~under nitrogen atmosphere  and (b)~in ambient air at a fixed bias voltage of \unit[5]{V}. The photoresponse was measured while switching on/off  the UV illumination source.}%
\label{fig:Transienten}
\end{figure}

In contrast, when exposed to ambient air~(Figure~\ref{fig:IV_NC}b), the dark $I$--$V$ characteristics exhibit strong super-linear behavior even at low voltage, indicating a much lower density of mobile charge carriers compared to inert conditions. At higher voltage, the scaling exponents clearly exceed the limit for trap-free SCLC conduction, which further suggests the presence of deep-level trap states. Upon UV illumination, the $I$--$V$ characteristics shift to almost Ohmic conduction, however, the photoconductivity upon air exposure is found to be only \unit[$5.8 \times 10^{-3}$]{S/m}~(\unit[$2.9 \times 10^{-4}$]{S/m}) for the rod-like~(spherical) particles, which is a decrease by two~(three) orders of magnitude compared to inert conditions.

Furthermore, we investigated the transient behavior of the photocurrents by switching on/off the UV illumination, and the results are depicted in Figure~\ref{fig:Transienten}. After having switched on the illumination, the photocurrent initially increases quickly, followed by a much slower growth process, until saturation has reached. Accordingly, a transient photocurrent decay can be observed when the illumination is switched off, and the decay time strongly depends on the surrounding gas atmosphere. Under nitrogen atmosphere~(Figure~\ref{fig:Transienten}a), the photoconductivity is found to be persistent on the time scale of hours, whereas it decays relatively fast within a few seconds to the dark conductivity level when the sample is exposed to ambient air~(Figure~\ref{fig:Transienten}b). In both cases, the values of the decay time are significantly smaller in case of the smaller spherical particles with the high surface-area-to-volume ratio.

We also studied the spectral photoresponse of the Al/ZnO/Al devices in ambient air~(see Supplemental Material, Figure~S7). From that analysis, we find that only excitation with photon energies above the band gap energy led to a notable increase of the conductivity compared to dark conditions. This finding indicates that electronic states within the band gap, e.g., populated surface states, have a low optical cross section and is in good agreement with the UV--visible absorption measurements on colloidal solutions. 

Our results regarding the transport mechanisms in ZnO nanoparticle films correspond well with the oxygen adsorption--desorption process as described above~(Figure~\ref{fig:adsorption_desorption}). Upon air exposure, the ZnO particles are depleted of mobile carriers, leading to strong injection-limited behavior in the dark. During UV illumination, electron-hole pairs are generated and become subsequently separated, as the holes are forced to the surface along the upward band bending, where they are likely to recombine with surface-trapped electrons. At the same time, the recombination probability for the remaining excess electrons in the conduction band decreases, giving rise to the enhanced~(Ohmic) photoconductivity. However, the photoconductivity in ambient air is several orders of magnitude lower than that under inert conditions, where the probability for readsorption is dramatically lower.

Regarding the different kind of ZnO nanoparticles used in this study, our results clearly suggest that carrier depletion related to surface adsorption becomes more important with increasing surface-area-to-volume ratio. This is mainly supported by the fact that the electrical properties of the smaller spherical particles are more severely affected by the presence of ambient oxygen compared to the larger rod-like ones, particularly concerning (i) the major difference between the photoconductivity under inert and ambient conditions and (ii) the the faster photocurrent decay after switching off the UV illumination.

\subsection{Ambipolar Charge Transport Properties of the Solar Cells}
\begin{figure}
\includegraphics{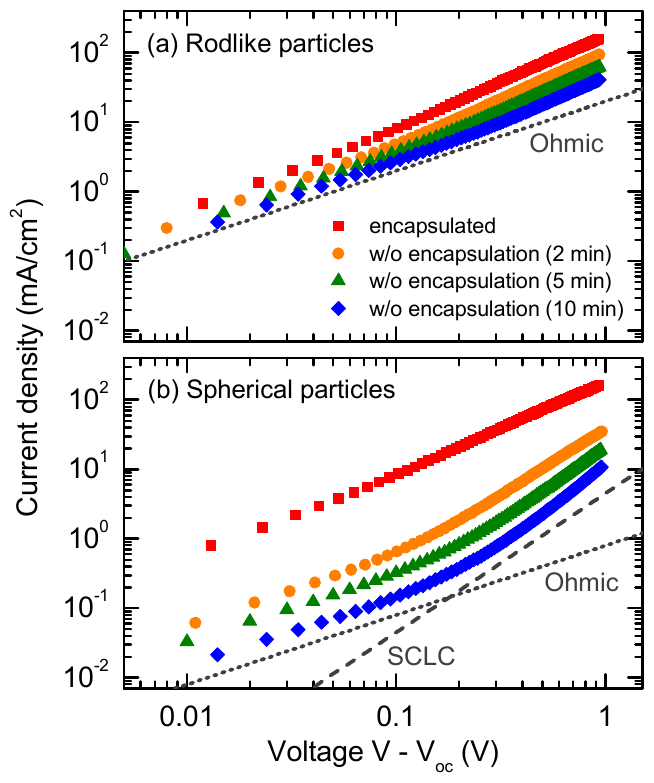}%
\caption{Log-log scaled $J$--$V$ characteristics under UV cut-off illumination for solar cells with interlayers processed from the (a)~rod-like and (b)~spherically shaped ZnO nanoparticles for varying exposure time to ambient air and after encapuslation. The voltage axis was shifted by $V_\text{oc}$. Dotted and dashed lines indicate the slopes expected for Ohmic and SCLC conduction, respectively.}%
\label{fig:IV_loglog}
\end{figure}

Accordingly, we investigated the ambipolar charge transport properties of the solar cells in more detail. Figure~\ref{fig:IV_loglog} shows log-log representations of the $J$--$V$ characteristics under UV cut-off illumination for non-encapsulated devices after different storage in ambient air, as well as for encapsulated devices. Herein, $V_\text{oc}$ was set to the origin of the voltage axis. For the larger rod-like particles~(Figure~\ref{fig:IV_loglog}a), it can be seen that the scaling behavior does not depend whether the cell is encapsulated or not. In all cases presented, the scaling exponent obtained via power-law fitting has a value of $\alpha = 1.3$, which indicates that the dominant charge transport mechanism does not change, at least, during the time scale of the experiment. This further justifies that the degradation of the $J$--$V$ curves upon air exposure mainly relates to increasing Ohmic losses at the ZnO/bulk heterojunction interface in case of the rod-like particles with the low surface-area-to-volume ratio, effectively giving rise to the increasing series resistance of the devices.

In contrast, in case of the smaller spherical particles~(Figure~\ref{fig:IV_loglog}b), the slope of the $J$--$V$ curves for the devices without encapsulation is strongly increased compared to the encapsulated devices. Without encapsulation, the scaling exponent increases with longer air exposure from $\alpha = 1.9$~(\unit[2]{min} air exposure) to $\alpha = 2.4$~(\unit[10]{min} air exposure), and, hence, approaches, even exceeds the threshold for SCLC conduction, whereas similar behavior to the case of the rod-like particles is obtained after encapsulation~($\alpha = 1.3$). In view of this, the S-shaped deformation in ambient air can be interpreted as a transition from an ohmic regime~($J \propto V$) at low voltages to an SCLC regime~($J \propto V^2$) at higher voltages, in agreement with the work of Wagenpfahl et al.\cite{Wagenpfahl2010}

The latter is also consistent with the picture of an effectively reduced surface recombination velocity caused by the formation of a charge extraction barrier. We suppose that the electron depletion in the nanocrystalline ZnO layer and the subsequent upward band bending resulting from the chemisorption of ambient oxygen species lead to an energetic barrier for electron transport and extraction at the ZnO/bulk heterojunction interface. Our findings suggest that the effective height of the energetic barrier correlates with the internal surface area of the nanocrystalline ZnO film. A lower internal surface area~(larger rod-like particles) leads to a moderate barrier height, mainly contributing to increased Ohmic losses, whereas, for a higher amount of internal surface area~(smaller spherical particles), electron depletion becomes more dominant, leading to an increased extraction barrier. Under extraction conditions~($V<V_\text{oc}$), the higher energetic barrier then results in the accumulation of majority charge carriers in the vicinity of the ZnO interface. Under injection conditions~($V>V_\text{oc}$), also less electrons are injected into the polymer--fullerene blend, leading to a lower bulk recombination rate and, consequently, an accumulation of holes at the anode. Both effects give rise to the formation of space charge and, thus, the transition to SCLC type conduction, causing the S-shape deformation of the $J$--$V$ curve. Under the presence of UV light, however, the extraction barrier effectively decreases, as the oxygen desorption becomes favorable over the adsorption process. This finding corresponds well to the beneficial effect of light-soaking, frequently observed for organic solar cells comprising ZnO interfacial layers.\cite{Chen2012,Lilliedal2010,Manor2012,Morfa2014,Cowan2014}


\section{Summary and Conclusions}
To summarize the main results, we have studied the impact of the surface-area-to-volume ratio of colloidal ZnO nanoparticles used as electron-selective interlayer in inverted organic solar cells on the electrical properties of the devices. Non-encapsulated devices, where oxygen can penetrate into the solar cells through the gas-permeable PEDOT:PSS top electrode, exhibited fast degradation of the dark current in forward direction on a time scale of minutes, the effect being more pronounced for small ZnO nanocrystals as compared to larger nanorods with a reduced surface-area-to-volume ratio. Under illumination, the injection current strongly increased, but only if the light source contained UV-light which can be absorbed in the ZnO phase. Qualitatively, the decreasing injection current can be explained in terms of oxygen adsorption at the nanoparticle surface which leads to band bending and the formation of a depletion zone at the ZnO surface. For the rod-like particles, the impact of this phenomenon was found to result in an increase of the series resistance of the devices. In contrast, for the smaller particles with a larger specific surface area, the degradation could not be explained simply by Ohmic losses. 

Therefore, we had a closer look at the transport properties and determined the photoconductivity of pure ZnO layers. The conductivity under illumination with UV light and in the dark differed by about three orders of magnitude, the difference being again larger for the smaller ZnO nanoparticles. In the dark, exposure to ambient air resulted in the occurrence of space-charge limited currents, but under UV illumination, the transport became of Ohmic nature. Concerning the ambipolar transport in illuminated solar cells, Ohmic behavior was found for solar cells with the larger rod-like particles. In contrast, exposure of non-encapsulated devices with the small nanocrystals led to a transition from Ohmic conduction to an SCLC regime on a time scale of minutes, if the UV part of the spectrum was removed.

From the detailed analysis and systematic variations of air exposure and illumination conditions, we could finally deduce the following scenario: If oxygen can penetrate the devices, adsorption at the ZnO surface leads to a depletion zone and band bending, which in turn results in the formation of a barrier for the injection and extraction of electrons. Under UV illumination, mobile carriers are generated in the ZnO phase, and adsorbed oxygen species can be removed from the surface so that the barrier height is reduced. Since the surface of the ZnO nanoparticles is involved into these mechanisms, it is comprehensible that the effects are more strongly pronounced in the case of smaller nanoparticles with larger surface-area-to-volume ratio. As a finding of this study, we observed in more detail that the specific surface area can be crucial for the nature of charge transport. With the small ZnO nanocrystals, a transition to a space-charge limited regime was observed under air exposure in the absence of UV light. This, in turn, resulted in the formation of an S-shaped $J$--$V$ curve under these conditions. In contrast, devices made with the rod-shaped nanoparticles showed a degradation effect as well, but it was much less pronounced, and S-shaped $J$--$V$ curves were not obtained, at least on the time scale investigated.

Finally, we have demonstrated that at least in case of using gas-permeable top electrodes like PEDOT:PSS films, light-soaking could only partly recover the initial $J$--$V$ behavior of the solar cells obtained directly after fabrication under inert atmosphere. Instead, we highlight the importance of a proper encapsulation of the devices from the ambient atmosphere, not only in order to protect the polymer--fullerene absorber against degradation, but also to preserve the favorable electron extraction capability of nanocrystalline ZnO layers on a long-term scale. With an involved UV curing step, the encapsulation procedure has also been successfully confirmed in ambient air.


\begin{acknowledgements}
The authors thank Janet Neerken, Dorothea Scheunemann, Manuela Schiek, Thomas Madena, and Matthias Macke for valuable support as well as the DYMAX Europe GmbH, Wiesbaden for providing the adhesive used for encapsulation. Funding from the EWE-Nachwuchsgruppe ``D\"{u}nnschichtphotovoltaik'' by the EWE AG, Oldenburg is gratefully acknowledged.
\end{acknowledgements}

\bibliography{ms}

\end{document}


\title{Supplemental Material for ``The Role of Oxygen Adsorption in Nanocrystalline ZnO Interfacial Layers for Polymer--Fullerene Bulk Heterojunction Solar Cells''}

\author{Sebastian Wilken}
\email{sebastian.wilken@uol.de}
\author{J\"urgen Parisi}
\author{Holger Borchert}
\affiliation{Institute of Physics, Energy and Semiconductor Research Laboratory, Carl von Ossietzky University of Oldenburg, 26111 Oldenburg, Germany}

\maketitle
\tableofcontents

\section*{X-Ray Diffraction}
\begin{figure}[h!]
\includegraphics{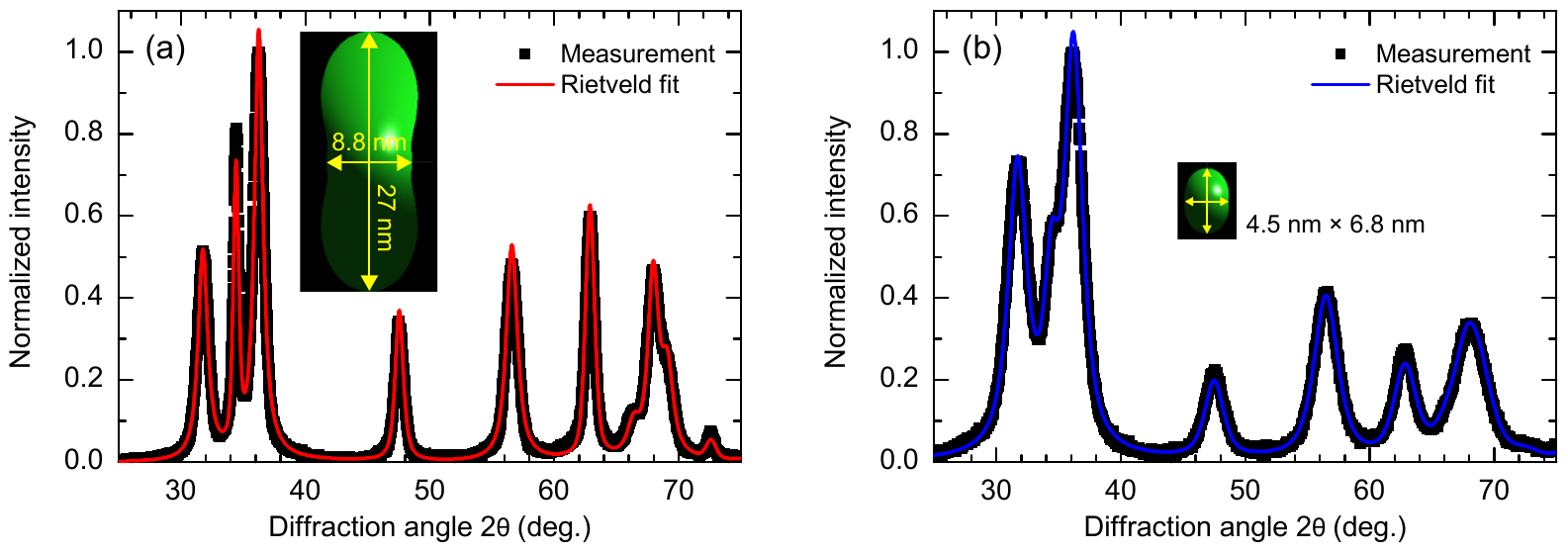}%
\caption{Powder X-ray diffraction patterns of the rod-like (a) and ``spherical'' (b) ZnO nanoparticles. The experimentally obtained data (squares) was fitted using Rietveld refinement (solid lines). For the representation in the figure, the background has been subtracted, and the data is normalized to the maximum intensity of the experimental data. The insets show the shape models obtained from the Rietveld analysis, represented at the same scale.}
\label{fig:xrd}
\end{figure}

Powder X-ray diffraction patterns of the ZnO nanoparticles, measured with a PANalytical X'PertPro MPD diffractometer, were analyzed by Rietveld refinement with the help of the software MAUD.\cite{Lutterotti2004} To simulate the microstructure of the elongated nanocrystals as well as preferred orientation effects, the size-strain model and the harmonic texture model developed by Popa were used.\cite{Popa1992,Popa1998} More details about the analysis procedure can be found in an earlier publication.\cite{Wilken2012} Figure \ref{fig:xrd} shows the experimental diffraction patterns of the two ZnO samples together with Rietveld fits. As a particularity of the size-strain model, an anisotropic crystallite shape can be modeled. From this analysis, we derived for the rod-like sample an average length of the rods of about \unit[27]{nm} along the c-axis of the hexagonal unit cell and an average width of \unit[8.8]{nm} perpendicular to the c-axis. In case of the smaller nanocrystals, the average sizes were \unit[6.8]{nm} and \unit[4.5]{nm} along and perpendicular to the c-axis, respectively. These shapes are visualized in Figure \ref{fig:xrd} as well. According to the TEM images shown in the main article, the result for the smaller particles may be understood as the average for a mixture of quasi-spherical particles and short rods composed of two attached spheres.

\clearpage
\section*{UV--Visible Absorption and Determination of the Optical Band Gap Energy}
\begin{figure}[h!]
\includegraphics{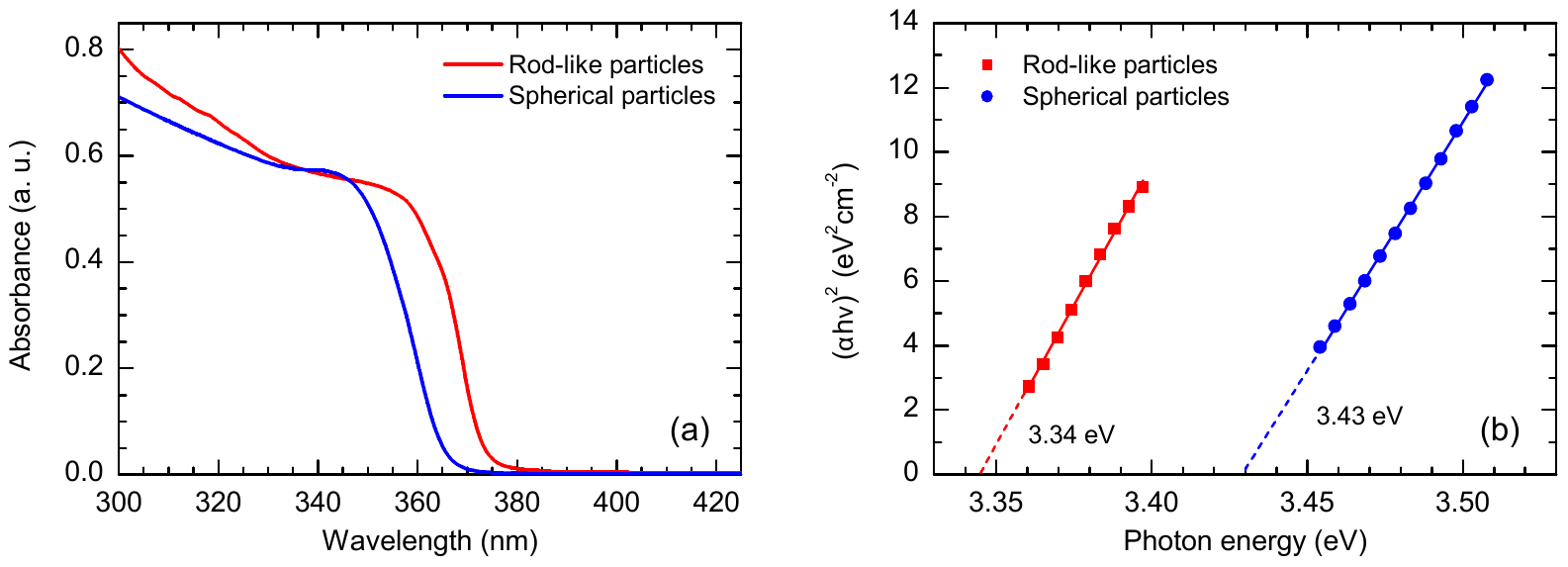}%
\caption{(a) UV-visible absorption of the rod-like and spherical ZnO particles in solution. (b) Tauc plots of the absorbance data. The values of the optical band gap energy given in the figure were obtained by extrapolating the linear part of the data to $(\alpha h \nu)^2 = 0$.}
\label{fig.Absorption}
\end{figure}

Figure \ref{fig.Absorption}a shows UV-visible absorption spectra for colloidal solutions of the rod-like and spherical ZnO particles in chloroform/methanol (9:1 v:v). The absorbance ($A$) was obtained by measuring the transmittance ($T$) of optically thin samples with a spectrophotometer (Varian Cary 100) using \unit[1]{cm} thick quartz cuvettes and calculating $A = -\log_{10} T$. Figure \ref{fig.Absorption}b shows Tauc plots corresponding to the absorption spectra to determine the optical band gap energy.\cite{Tauc1966} For direct semiconductors, the absorption coefficient $\alpha$ is related to the photon energy $h\nu$ according to
\begin{equation}
\alpha h \nu \approx A^\ast (h \nu - E_\text{g})^{1/2},
\end{equation}
where $E_\text{g}$ is the band gap energy and $A^\ast$ a frequency-independent constant. Hence, $E_\text{g}$ can be estimated from experimental absorption data by plotting $(\alpha h \nu)^2$ versus $h \nu$ and extrapolating the linear part of the data to $(\alpha h \nu)^2 = 0$. According to the Beer--Lamber law, the absorption coefficient was calculated by $\alpha = (A \cdot \ln 10)/d$, where $d$ is the optical path length, herein determined by the cuvette thickness.

\clearpage
\section*{Surface Morphology of the ZnO Films}
\begin{figure}[h!]
\includegraphics{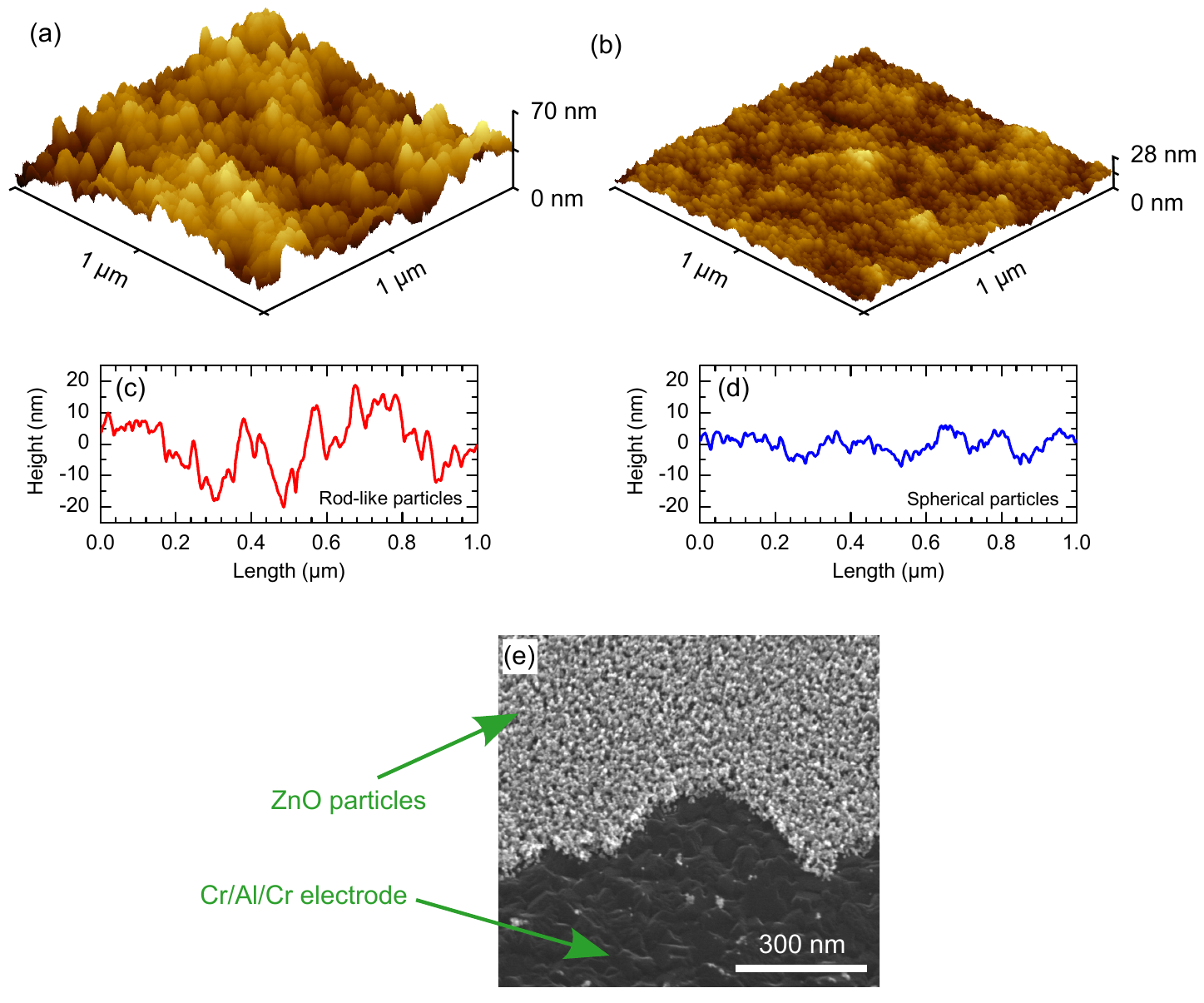}%
\caption{Surface morphology of thin films made from the rod-like (a) and spherical (b) ZnO particles measured by AFM. (c,d) Typical height profiles extracted from the AFM scans. (e) SEM image of a ZnO nanoparticle film (spherical particles) processed onto a Cr/Al/Cr electrode.}
\label{fig.AFM}
\end{figure}

To investigate the surface morphology of the ZnO films, ZnO particles were processed onto Cr/Al/Cr electrodes under the same conditions as for solar cell fabrication, i.e., spin-coating under inert atmosphere at \unit[1500]{rpm} without any further treatment. The morphology was then measured using an atomic force microscope (AFM, Agilent 5420) in intermittent-contact mode (see Figure \ref{fig.AFM}). The root mean square roughness of the films (see main text) was calculated from $\unit[(1 \times 1)]{\mu m^2}$ scans with the software \texttt{Gwyddion}. To visualize the thin film quality, Figure \ref{fig.AFM}e exemplarily shows a scanning electron microscopy (SEM, FEI Helios NanoLab 600i) image of a nanocrystalline ZnO film made from the spherical particles.

\clearpage
\section*{Additional $J$--$V$ Characteristics of the Solar Cells}
The $J$--$V$ curves are averaged over six individual devices in each case.

\begin{figure}[h!]
\includegraphics{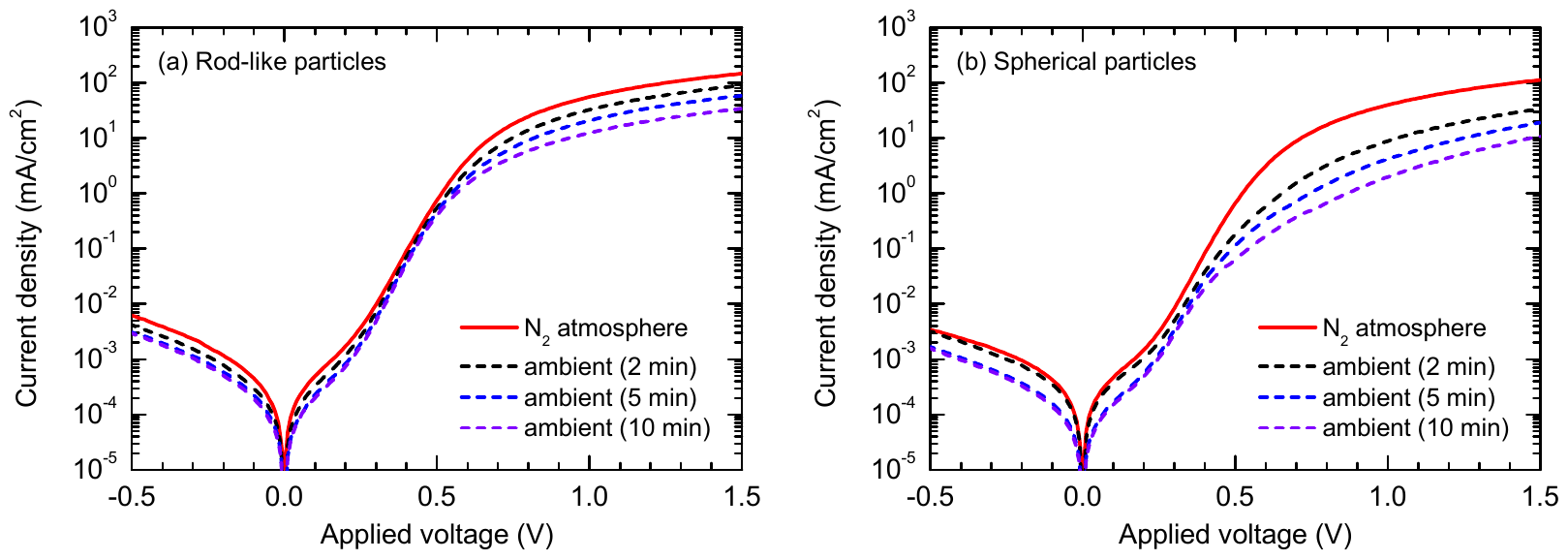}%
\caption{Averaged dark $J$--$V$ curves for non-encapsulated solar cells with interlayers made from the rod-like (a) and spherical (b) ZnO particles, recorded initially under nitrogen atmosphere and, subsequently, after 2, 5, and \unit[10]{min} storage in ambient air.}
\label{fig.IV_dark}
\end{figure}

\begin{figure}[h!]
\includegraphics{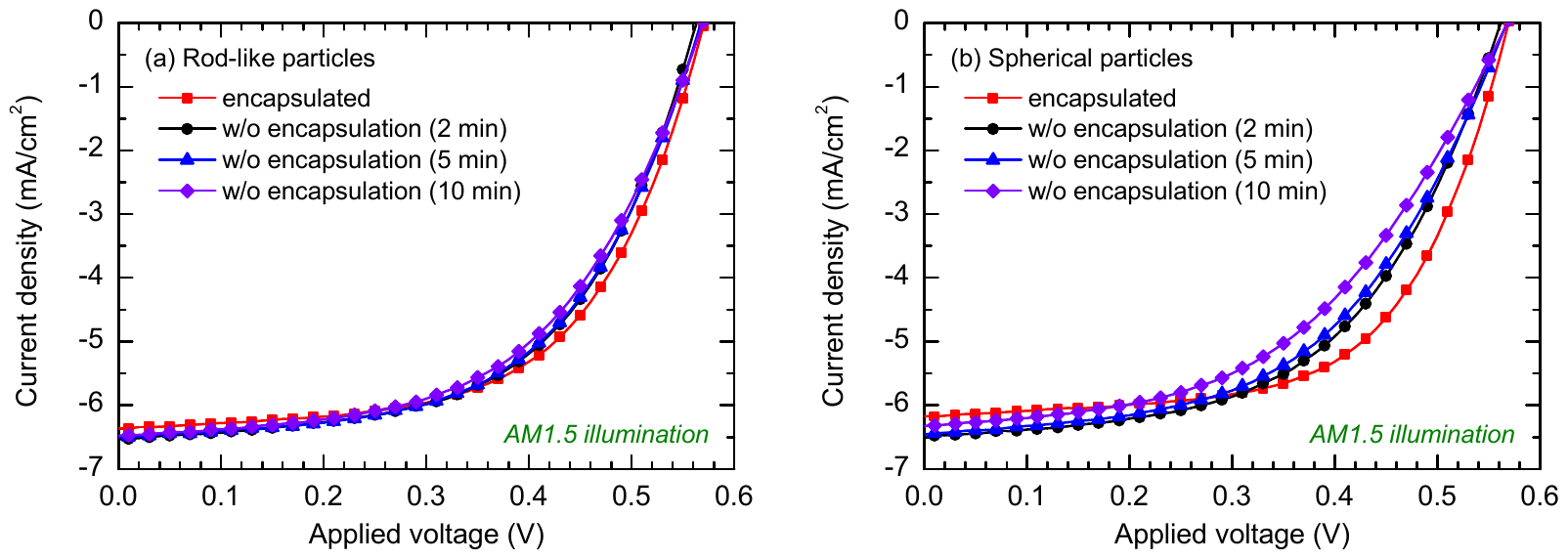}%
\caption{Averaged $J$--$V$ characteristics under standard AM1.5 illumination for solar cells with interlayers made from the rod-like (a) and spherical (b) ZnO particles, recorded initially without encapsulation after 2, 5, and \unit[10]{min} storage in ambient air and, subsequently, after having been encapsulated.}
\label{fig.IV_light}
\end{figure}

\clearpage
\section*{Stability of the Encapsulated Solar Cells}
\begin{figure}[h!]
\includegraphics{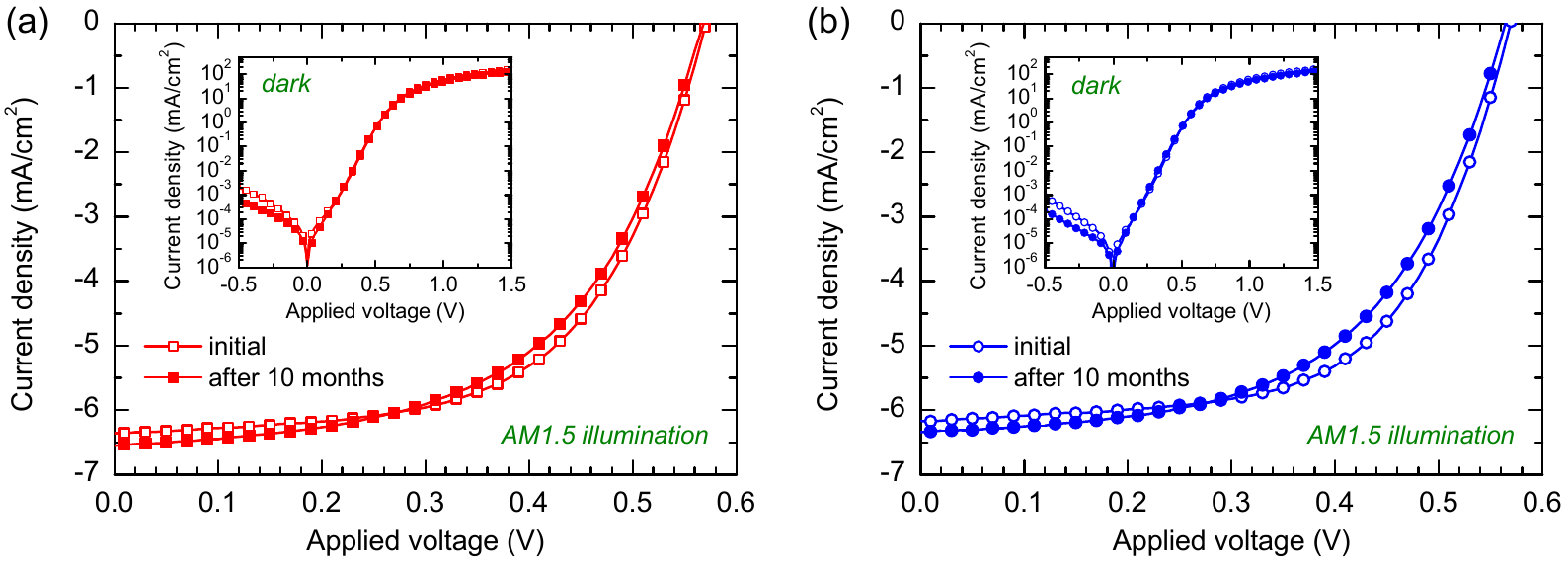}%
\caption{$J$--$V$ characteristics in the dark (insets) and under AM1.5 illumination for encapsulated solar cells with interlayers made from the rod-like (a) and spherical (b) ZnO particles, recorded immediately after the encapsulation procedure and after a storage period of \unit[10]{months}.}
\label{fig.stability}
\end{figure}

To demonstrate the inherent stability of the encapsulated devices, Figure \ref{fig.stability} and Table \ref{tab.stability} present the photovoltaic performance obtained (i) immediately after the encapsulation procedure, and (ii) after \unit[10]{months} storage under dark conditions in a nitrogen glove box. After storage, approximately 95\% of the initial power conversion efficiency is preserved, which indicates the high suitability of the applied encapsulation method. The apparent variations in $J_\text{sc}$ and FF are supposed to be mainly related to reorganisation effects in the polymer--fullerene blend rather than inferior charge injection/extraction properties at the ZnO/bulk heterojunction interface, as is evidenced by the fact that the magnitudes of the dark current density under forward bias (see insets in Figure \ref{fig.stability}) are nearly the same before and after the storage period.

\begin{table}[h!]
  \caption{Average photovoltaic performance parameters of the encapsulated solar cells, obtained immediately after the encapsulation procedure as well as after \unit[10]{months} storage.}
  \begin{ruledtabular}
  \begin{tabular}{lllll}
   \toprule
    Particles & Parameter & Initial & After \unit[10]{months} & Percentage change\\
    \midrule
   	rod-like & $V_\text{oc}$ & \unit[570]{mV} & \unit[568]{mV} & $-0.5\%$\\
   					 & $J_\text{sc}$ & \unit[6.36]{mA/cm$^2$} & \unit[6.55]{mA/cm$^2$} & $+2.9\%$\\
   					 & FF & 59.0\% & 54.9\% & $-7.0\%$\\
   					 & PCE & 2.14\% & 2.04\% & $-4.6\%$\\
					 \\
   	spherical & $V_\text{oc}$ & \unit[570]{mV} & \unit[565]{mV} & $-0.9\%$\\
   					 & $J_\text{sc}$ & \unit[6.18]{mA/cm$^2$} & \unit[6.34]{mA/cm$^2$} & $+2.7\%$\\
   					 & FF & 60.7\% & 55.6\% & $-8.4\%$\\
   					 & PCE & 2.14\% & 1.99\% & $-6.9\%$\\
  \bottomrule
  \end{tabular}
  \end{ruledtabular}
  \label{tab.stability}
\end{table}

\clearpage
\section*{Spectral Photoresponsivity of the ZnO Films}
For the spectrally resolved photoresponsivity measurements, the Al/ZnO/Al devices were illuminated in ambient air with monochromatic light, provided by a \unit[75]{W} xenon short arc lamp combined with a monochromator (Bentham Instruments TMc300). The current through the ZnO sample was monitored with a source measurement unit (Keithley 2400) at a fixed bias voltage of \unit[+50]{V}. The apparent photocurrent ($I_\text{ph,ZnO}$) was then calculated by subtracting the dark current from the current obtained under illumination. To calibrate the optical system, we also measured the photocurrent of a Si photodiode ($I_\text{ph,Si}$) with a known spectral responsivity ($\text{SR}_\text{Si}$) at zero bias. The spectral photoresponsivity of the ZnO devices, $\text{SR}_\text{ZnO}$, was then determined by
\begin{equation}
\text{SR}_\text{ZnO} = \text{SR}_\text{Si} \cdot \frac{I_\text{ph,ZnO}}{I_\text{ph,Si}}.
\end{equation}

\begin{figure}[h!]
\includegraphics{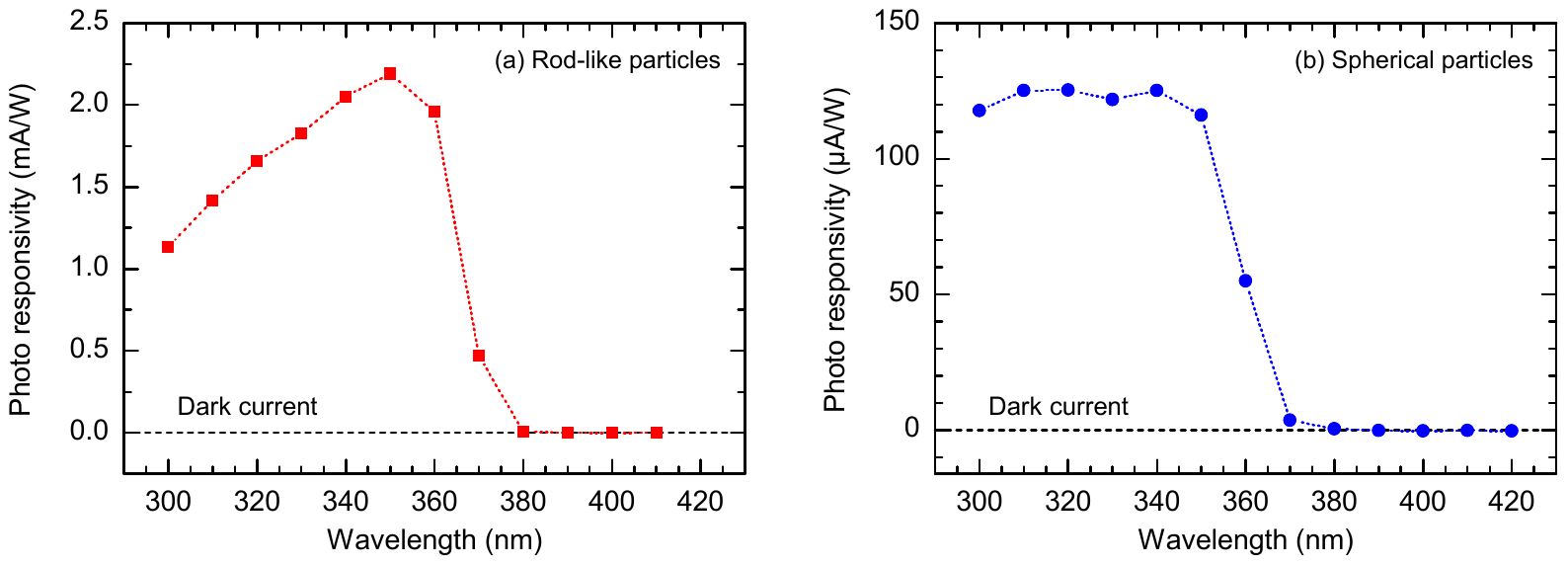}%
\caption{Spectrally resolved photoresponsivity in ambient air of thin films made from the rod-like (a) and spherical (b) ZnO particles employed in a lateral Al/ZnO/Al structure.}
\label{fig.SR}
\end{figure}

Figure \ref{fig.SR} shows the resulting photoresponsivity spectra for the rod-like and spherically shaped ZnO nanoparticles. Within the resolution of the setup, no photoconductivity upon excitation with photon energies below the band gap energy could be observed in both cases. Note the different scaling of the ordinate (\unit{mA/W} vs. \unit{$\mu$A/W}).

\bibliography{supplement}